\documentclass[11pt]{article}
\usepackage{epsfig}

\begin{document}

\noindent
{\huge Atmospheric neutrinos}

\noindent
\hspace{1cm}{\Large in the context of muon and neutrino radiography}

\vspace{.5cm}

\begin{raggedright}  

{\it Thomas K. Gaisser\\
Bartol Research Institute and Department of Physics and Astronomy\\
University of Delaware\\
Newark, DE 19716}
\end{raggedright}

\section*{Abstract}

{\small Using the atmospheric neutrinos to probe the density
profile of the Earth depends on knowing the angular distribution of the neutrinos 
at production and the neutrino cross section.  This paper reviews the essential
features of the angular distribution with emphasis on the relative contributions
of pions, kaons and charm.}

\section{Introduction}

In principle atmospheric neutrinos can be used to measure the density profile of
the Earth (integrated along a chord) by comparing the observed rate of upward-moving neutrino-induced muons
as a function of zenith angle with what is produced in the atmosphere.
The measurement depends on several factors: the zenith angle dependence of neutrino production 
in the atmosphere, the differential neutrino cross section as a function of energy, properties of
muon propagation in the medium surrounding the detector and the angular and energy
resolution of the detector.  Gonzalez-Garcia, Halzen and Maltoni [2008] show that 
it may be possible to measure some aspects of the core-mantle transition region with the
IceCube detector currently under construction at the South Pole [Karle et al., 2007].
A data accumulation of order ten years or more would be needed.  

The critical energy range for such a measurement is $10\le E_\nu\le 100$ TeV where the
absorption of the Earth becomes important, first for neutrinos coming straight up
through the diameter of the Earth, and at higher energy for larger angles.
Hoshina [2008] discusses the angular resolution of IceCube for neutrino-induced muons
from muon neutrinos in this energy band, where more than 50\% of events can be
reconstructed within one degree of their true direction.  The charged current neutrino
cross sections in this energy range are known to within $\pm2$\% 
[Cooper-Sarkar and Sarkar, 2008].
In this paper, I focus on the angular distributions of muon neutrinos produced in the
atmosphere and their uncertainties.

\section{Angular dependence of atmospheric neutrinos}

When cosmic ray protons and nuclei enter the atmosphere they interact and produce 
cascades of secondaries, including charged hadrons, some of which decay to produce
muons and neutrinos.  Muon neutrinos of high energy are produced primarily in the decay chain
\begin{eqnarray}
\label{chain}
K^\pm\rightarrow & \mu^\pm & +\; \nu_\mu(\overline{\nu}_\mu) \nonumber   \\
                   &\;\;\;\;\searrow &					        \\
 &  &e^\pm + \nu_e(\overline{\nu}_e)+\overline{\nu}_\mu(\nu_\mu),\nonumber
\end{eqnarray} 
There are similar expressions for decay chains initiated by pions and charmed hadrons.
In the energy range of interest here, decay of muons is rare and its contribution
to muon neutrinos can be neglected.  Moreover, the contribution from kaons is large
($\sim 80$\%) as compared to the contribution from decay of charged pions ($\sim 20$\%).
The contribution from charm decay is small, but may become significant in the energy
region of interest here, as discussed  below.

Neutrino fluxes are determined by a combination of four factors:
\begin{enumerate}
\item the energy spectrum and composition of the primary cosmic radiation,
\item production of pions, kaons and charmed hadrons in collisions of nucleons
and other hadrons with nuclei of the atmosphere,
\item kinematics of hadron decays to neutrinos, and
\item the density profile of the atmosphere.
\end{enumerate}
The primary spectrum can be approximated by a sequence of
power laws.  The energy range relevant for $10$-$100$~TeV
neutrinos is of order $30$~TeV to $3$~PeV, where the
spectrum of nucleons is approximately
\begin{equation}
\label{primary}
{{\rm d}N\over {\rm d}\ln(E)}\,\,\propto\,\,E^{-\gamma},
\end{equation}
with $\gamma\approx 1.7$ [Gaisser \& Stanev, 2008].  

For each of the main neutrino parents ({\it i}), there is a critical energy,
\begin{equation}
\epsilon_i\,=\, m_ic^2\times\left({h_0\over c\tau_i}\right),
\end{equation}
where $h_0$ is the scale height of an exponential approximation to the
atmosphere and $m_i$ and $\tau_i$ are respectively the mass and rest
lifetime of a particle of type $i=\pi^\pm,K^\pm,D^\pm$.  When 
\begin{equation}
\label{secant}
E_\nu\,\,>\,\,\sec(\theta)\times\epsilon_i
\end{equation}
re-interaction of the parent hadron is favored over its decay.  Approximate
numerical values are given in Table~1.  

The energy dependence of the competition between decay and re-interaction of
the different parent mesons
affects the neutrino energy spectrum in an essential way.  These effects
can be displayed explicitly using analytic approximations for the
neutrino energy spectrum.  
In the power-law approximation
and at high energy, the spectrum of neutrinos,
$\phi_\nu$, is approximated well by
\begin{eqnarray}
\label{angular}
\phi_\nu(E_\nu)& = & \phi_N(E_\nu) \nonumber \\
 & \times & \left\{{A_{\pi\nu}\over 1 + 
B_{\pi\nu}\cos(\theta)E_\nu / \epsilon_\pi}
\,+\,{A_{K\nu}\over 1+B_{K\nu}\cos(\theta)E_\nu / \epsilon_K}\right.\nonumber \\
& & \left. +\,\,\,{A_{{\rm charm}\,\nu}\over 1+B_{{\rm charm}\,\nu}\cos(\theta)E_\nu / \epsilon_{\rm charm}}\right\},
\end{eqnarray}
where $\phi_N(E_\nu) = dN/d\ln(E_\nu)$ is the primary spectrum
of nucleons ($N$) evaluated at the energy of the neutrino [Gaisser, 1990].  The factors $A_{i\nu}$
contain the physics of meson production weighted by the spectrum and the decay kinematics
contained in Equation~\ref{chain}.  

As an example, $$
A_{\pi\nu}\,=\,{Z_{N\pi}\over 1\,-\,Z_{NN}}\;{(1-r_\pi)^\gamma\over \gamma\,+\,1}. $$
Here $$Z_{ab} = {1\over \sigma_{a}}\int_0^1\,x^\gamma {{\rm d}\sigma_{ab}
 (x)\over {\rm d}x}$$
is the spectrum weighted moment for the interaction process in which a particle $a$
interacts with a nucleus in the atmosphere and produces a secondary particle $b$
that has a fraction $x$ of the lab energy of the projectile.  The factor involving
$r_\pi = m_\mu^2/m_\pi^2$ is the spectrum weighted kinematic factor for
the decay $\pi\rightarrow \mu\,+\,\nu$ and $\gamma\approx 1.7$ is the integral
spectral index in a power-law approximation to the primary cosmic ray spectrum.

\begin{table}
\label{critical}
\caption{Critical energies in GeV}
\begin{center}
\begin{tabular}{lll} 
\hline
$\epsilon_\pi$ & $\epsilon_K$ & $\epsilon_{\rm charm}$ \\ \hline
$115$ & $850$ & $\sim 5\times 10^7$\\ 
\hline
\end{tabular}
\end{center}
\end{table}

At low energy, the neutrino spectrum follows the same
power law as the parent hadrons.  At high energy, the probability
of decay of a charged pion or kaon is suppressed relative
to hadronic interaction by a factor proportional to the energy of the meson.
This suppression is represented by the denominator of each term in
Eq.~\ref{angular}.  Asymptotically at high energy the neutrino spectrum
has an extra factor of $1/E$ relative to the spectrum at low energy,
which is proportional to the primary cosmic-ray spectrum.
 This steepening of the neutrino spectrum occurs at higher energy for larger
zenith angles as a consequence of the
$\sec(\theta)$ factor, which occurs because particle production occurs
higher in less dense atmosphere for large angles. The factors $B_{i\nu}$ 
in the denominators are ratios of spectrum-weighted
kinematic factors for meson decay to neutrinos to account for the
fact that the decay occurs from a steeper spectrum at higher energy.    
Close to the horizontal
($\theta > 70^\circ$) the curvature of the Earth is significant and
the secant is to be evaluated at the local zenith angle where
the particle production occurs, which is less than the zenith angle
of the trajectory at the detector~[Lipari, 1993]. 

 Parameterizations of the form of Eq.~\ref{angular} represent detailed
numerical and Monte Carlo calculations rather well.  The corresponding
parametric equations for muons can be compared directly to the many
measurements of atmospheric muons.  Such a comparison is made in Fig.~24.4
Gaisser \& Stanev [2008], for example.  (See also Fig.~4 of Gaisser [2005]).

Uncertainties in the angular dependence arise primarily from uncertainties in the level
of kaon production for $E_\nu\sim 10$~TeV
and at higher energy from the more uncertain level of charm production.
Ideally one would simply measure the downward atmospheric neutrino flux
and compare it with the upward neutrino flux from the opposite direction
and use the ratio Earth-in to Earth-out to measure the density profile
in a way that is independent of the intrinsic angular dependence of the
atmospheric neutrino beam.  In reality this unfortunately cannot be done because
the relatively high intensity of downward atmospheric muons hides the
downward atmospheric neutrinos.   Because of the close genetic relation
between $\nu_\mu$ and $\mu$ of Eq.~\ref{chain} and the corresponding 
equation for pions, a measurement of downward atmospheric muons provides
some control of the angular dependence.  This constraint, while valid
and worth pursuing, is of limited practical use in this context
because of the fact that most muons come from decay of charged pions
while, in the high energy range of interest here, 
most neutrinos come from decay of kaons.

\begin{figure*}[t]
\centerline{\includegraphics[width=12cm,clip]{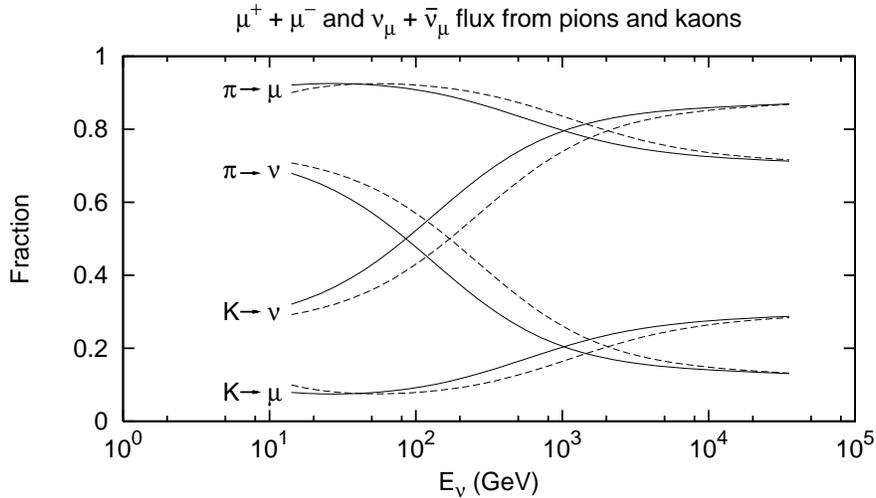}} 
\label{K2pi}
\caption{Fraction of muons and muon neutrinos from pion decay 
and from kaon decay vs. neutrino energy.
Solid lines for vertical, dashed lines for zenith angle $60^\circ$.}
\end{figure*}

\section{Neutrinos from kaons}
Each term in Eq.~\ref{angular} exhibits the same characteristic
dependence on energy and zenith angle.  Kaons become increasingly
important with increasing energy because $\epsilon_K\,>\,\epsilon_\pi$.
This means that the contribution from kaons retains a spectrum close
to that of the primary cosmic rays while the pion contribution is already 
beginning to steepen.  The location of this transition depends on the
$K/\pi$ ratio at production and is of the order of $100$~GeV, as shown
in Fig.~1.   
It occurs at higher energy for larger zenith angle.
Note that kaons never become the dominant source of atmospheric muons.
This is largely a consequence of the fact that the muon mass is not much
less than its parent pion so the $\nu_\mu$ carries
only a small fraction of the energy in $\pi\rightarrow\mu\nu_\mu$.

Uncertainties in the calculated intensity of atmospheric neutrinos at
high energy due to uncertain knowledge of hadronic interactions
were estimated by Agrawal {\it et al.} [1996].  The single largest
source of uncertainty in the TeV range and above is from the
uncertainty in kaon production, in particular in the factor $Z_{pK^+}$.
This uncertainty was estimated as $\pm13$\% in the 10 - 100 TeV range.

The recent measurement of the $\mu^+\,/\,\mu^-$ ratio above $1$~TeV 
with the MINOS far detector [Adamson {\it et al.}, 2007] reflects the importance
of kaon production in an interesting way.  The charge ratio increases
as kaons become dominant, from which one can infer that the 
$K^+\,/\,K^-$ ratio is larger than the $\pi^+\,/\,\pi^-$ ratio.
It is likely that the specific process of associated production
\begin{equation}
p\,\,+\,\,{\rm air}\,\,\rightarrow\,\,K^+\,+\,\Lambda\,+ {\rm anything}
\label{associated}
\end{equation}
plays an important role here.  Adamson {\it et al.} have shown how their
measurement can constrain the parameters that describe the 
production of kaons relative to pions and their charge ratios.
Because of the genetic relation
between atmospheric $\mu$ and $\nu_\mu$, the MINOS measurement,
in combination with measurements of $K\,/\,\pi$ production at
high energy, has the potential also to make the prediction of the
angular distribution of atmospheric muon neutrinos more precise.

\section{Neutrinos from charm}

Although the charm channel for atmospheric neutrinos is formally the
same as for pions and kaons, there are two large quantitative differences.
One is that production of charm in hadronic interactions 
is very much smaller than production of pions and kaons.
The other is that the critical energy is so high (because of
the short charm lifetime) that 
muons and neutrinos from decay of charmed hadrons will continue
in the low-energy regime of Eq.~\ref{angular}
with the same, relatively hard, spectrum as the primary
cosmic-ray nucleons and with an isotropic angular
distribution.  

\begin{figure*}[t]
\centerline{\includegraphics[width=10cm,clip]{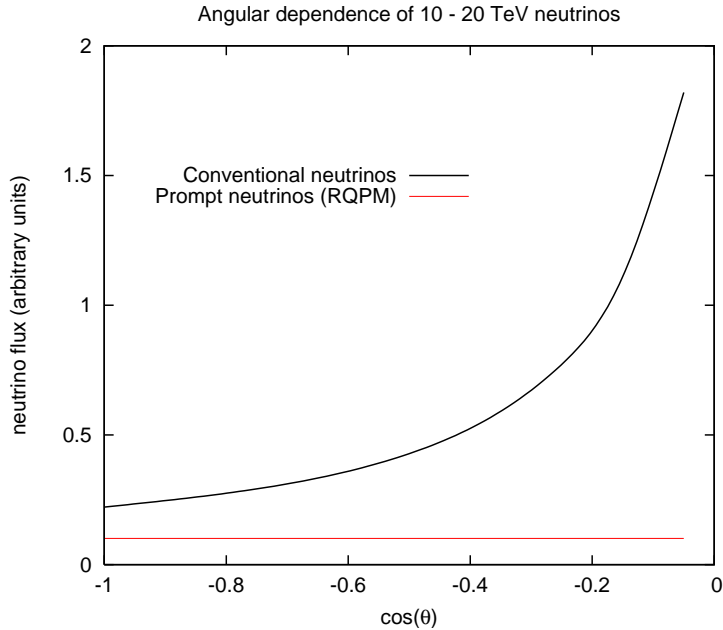}} 
\caption{The curved line shows the angular dependence of $\nu_\mu$ from
decay of pions and kaons while the horizontal line shows the isotropic
prompt component in the model of Bugaev et al. [1998] for $E_\nu = 10$
to $20$~TeV.}
\label{prompt}
\end{figure*}

Because of its harder spectrum, the
charmed contribution will
eventually become the dominant source of atmospheric $\nu_\mu$,
adding an isotropic ''prompt" component to the secant($\theta$)
dependence of the kaon and pion contributions. 
The transition energy is at some energy $E_\nu > 10 TeV$, which
therefore may be in the energy range relevant for Earth tomography
with neutrinos.  Because
the level of charm production is so uncertain, this is a significant
source of uncertainty in the angular distribution of the atmospheric
neutrino beam in this high energy range.  Gelmini, Gondolo \& Varieschi [2003]
assemble constraints from measurements of the angular
and energy distributions of atmospheric muons on charm production.
An example of a model for charm production that is near the upper limit
is the RQPM model of Bugaev {\it et al.} [1998].  Figure~\ref{prompt}
shows the angular distribution in the energy range $10 < E_\nu < 20$~TeV
using this model.  The prompt and normal components are shown separately.

In the model of Bugaev {\it et al.} atmospheric neutrinos from charm
decay become equal in intensity to neutrinos from decay of kaons and pions
at approximately $100$~TeV.
A more recent calculation [Enberg, Reno \& Sarcevic, 2008] predicts a level
of charm production an order of magnitude lower than the model of Bugaev {\it et al.}.
In this case, the crossover occurs about a factor of two higher in energy.

\section{Conclusion: detecting neutrinos}
The neutrino effective area is defined so that the product of neutrino intensity
multiplied by the effective area gives the event rate.  
The probability that a muon neutrino on a trajectory that will intercept the
detector gives a visible muon in the detector is
\begin{equation}
P(E_\nu,E_{\mu,{\rm min}})\,\,=\,\,N_A\,\int_{E_\mu,{\rm min}}^{E\nu}\,{\rm d}E_\mu
{{\rm d}\sigma_\nu\over {\rm d}E_\mu}\,R(E_\mu,E_{E_{\mu,{\rm min}}})\, ,
\label{nuprob}
\end{equation}
where the integrand is the product of the charged current differential cross section
and $R$ is the average distance traveled by a muon with energy $E_\mu$ at production
before its energy is reduced to $E_{\mu,{\rm min}}$, the minimum energy required
upon entering the detector for the event to be reconstructed well.  The additional 
contribution from neutrinos that interact within a large detector can be computed
in a straightforward way and added as an extra contribution.

\begin{figure*}[t]
\centerline{\includegraphics[width=12cm,clip]{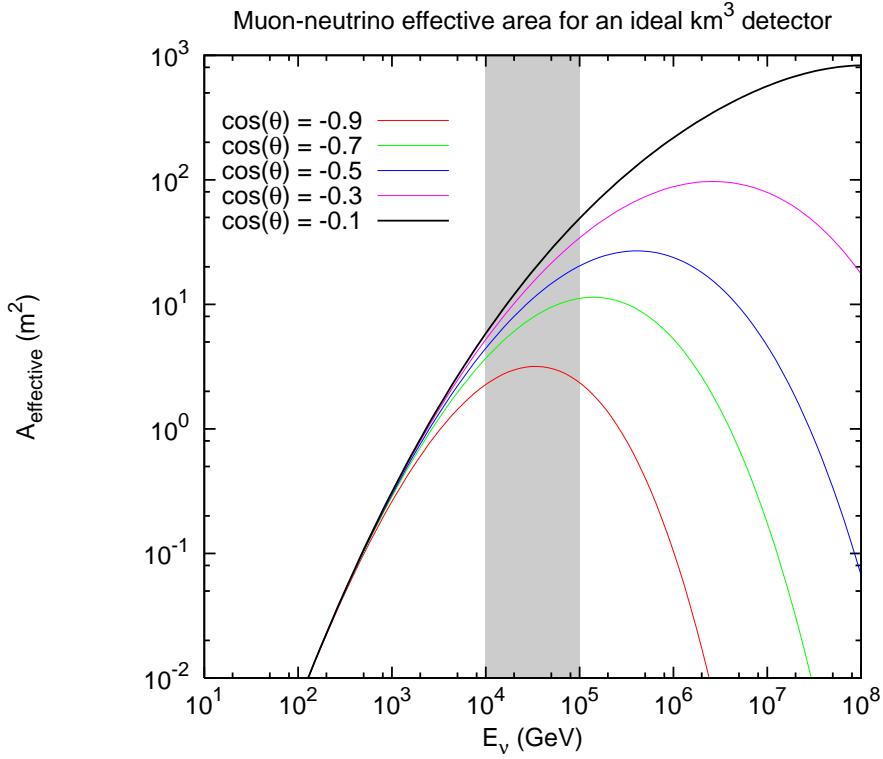}} 
\caption{Muon neutrino effective area for an idealized spherical
detector with cross-sectional area of $1$~km$^2$.}
\label{Atheta}
\end{figure*}

The neutrino effective area is
\begin{equation}
A_{\rm eff}(\theta,E_\nu)\;=\;\epsilon(\theta)\,A(\theta)\,P(E_\nu,E_{\mu,{\rm min}})
\,e^{-\sigma_\nu(E_\nu)N_A\,X(\theta)},
\label{Aeff}
\end{equation}
where $\epsilon(\theta)$ is the efficiency for a detector of projected area
$A(\theta)$ to detect a muon incident at zenith angle $\theta$.  The exponential
expresses the muon attenuation in the Earth for angle $\theta$ below
the horizon, where $X(\theta)$ is the amount of matter (g/cm$^2$) along
the chord through the Earth.
  Figure~\ref{Atheta} shows $A_{\rm eff}$ for an idealized
spherical detector with physical projected area of one km$^2$ for several
different zenith angles.  The shaded region indicates the energy range
of interest between 10 and 100 TeV.  Because the neutrino spectrum
decreases strongly with increasing energy, most of the rate will be
near the lower end of this range.  In Fig.~\ref{prompt}, the prompt
contribution is about 40\% of the normal $\pi$, K channel contribution
to the intensity of $\nu_\mu$ near $\cos(\theta)\approx -0.9$,
which roughly corresponds to the core region.
Because the level of charm is uncertain, this source of uncertainty
in the angular distribution of $\nu_\mu$ will be a significant limitation
to the measurement of the core-mantle transition.
Alternatively, a careful measurement of the angular distribution
of atmospheric neutrino induced muons with good energy resolution
in this energy range may clarify the level of charm production.

\subsection*{Acknowledgments}
This research is supported in part by the U.S. Department of Energy, DE-FG02-91ER40626.

\section*{References}

{\small
\noindent
P. Adamson {\it et al.}, MINOS Collaboration,
``Measurement of the atmospheric muon charge ratio at TeV energies 
with the MINOS detector", \textit{Phys. Rev. D}, \textbf{76}, 072005 (2007). 
\vspace{.1cm}

\noindent
V. Agrawal, T.K. Gaisser, P. Lipari \& T. Stanev, ``Atmospheric neutrino flux
above 1 GeV", \textit{Phys. Rev. D} \textbf{53}, 1314 (1996).
\vspace{.1cm}
  
\noindent
E. V. Bugaev, A. Misaki, V. A. Naumov, T. S. Sinegovskaya, S. I. Sinegovsky, \& N. Takahashi, ``Atmospheric muon flux at sea level, underground, and underwater",
\textit{Phys. Rev. D}, \textbf{58}, 054001 (1998).
\vspace{.1cm}

\noindent
A. Cooper-Sarkar and S. Sarkar, ''Predictions for high energy 
neutrino cross-sections from the ZEUS
global PDF fits", \textit{JHEP}, \textbf{01}, 075 (2008).
\vspace{.1cm}

\noindent
Enberg, R., M.H. Reno and I. Sarcevic,
``Prompt neutrino fluxes from atmospheric charm", \textit{Phys. Rev. D} \textbf{78}, 043005 (2008).
\vspace{.1cm}

\noindent
T.K. Gaisser ``Cosmic Rays and Particle Physics" (Cambridge University Press, 1990).  Japanese edition with K. Kobayakawa (Maruzen, 1997).
\vspace{.1cm}

\noindent
T.K. Gaisser ``Outstanding Problems in Particle Astrophysics", in \textit{Neutrinos and
Explosive Events in the Universe} (ed. M.M. Shapiro, T. Stanev \& J.P. Wefel, NATO Science
Series II, Mathematics, Physics and Chemistry - Vo. 209, Springer, 2005).  
astro-ph/0501195.
\vspace{.1cm}

\noindent
T.K. Gaisser \& Todor Stanev, ``Cosmic Rays", in \textit{Reviews of Particle Properties},
Claude Amsler, Michael Doser et al., [Particle Data Group], \textit{Phys. Lett. B} \textbf{667} (2008) 254-260.
\vspace{.1cm}

\noindent
G. Gelmini, P. Gondolo \& G. Varieschi, \textit{Phys. Rev. D} \textbf{67} (2003) 017301.
\vspace{.1cm}

\noindent
M.C. Gonzalez-Garcia, F. Halzen, M. Maltoni, H.K.M. Tanaka, ``Radiography of the Earth's Core and Mantle
with Atmospheric Neutrinos", \textit{Phys. Rev. Letters}, \textbf{100}, 061802 (2008).
\vspace{.1cm}

\noindent
K. Hoshina for the IceCube Collaboration, this volume.
\vspace{.1cm}

\noindent
A. Karle for the IceCube Collaboration, ``IceCube--construction, status, performance results
of the 22 string detector", to be published in Proc. 30th International Cosmic Ray Conference (Merida),
see arXiv:0711.0353v1 7--10 (2007).
\vspace{.1cm}

\noindent
P. Lipari, ``Lepton spectra in the earth's atmosphere",
 \textit{Astropart. Phys.} \textbf{1}, 195-227 (1993).
}

\end{document}